\documentclass[prb, twocolumn]{revtex4}
\usepackage{amsmath}
\usepackage{graphicx}
\usepackage{psfrag}

\begin{document}

\title{On the stability of electrostatic orbits}

\author{Shubho Banerjee}
\email{banerjees@rhodes.edu}
\affiliation{Rhodes College, Department of Physics, Memphis,
Tennessee 38112}

\author{Bradford Taylor}
\affiliation{Rhodes College, Department of Physics, Memphis,
Tennessee 38112}

\author{Anand Banerjee}
\affiliation{University of Maryland, Department of Physics, College Park,
Maryland 20742}


\begin{abstract}
We analyze the stability of two charged conducting spheres orbiting each other. Due to charge polarization, the electrostatic force between the two spheres deviates significantly from $1/r^2$ as they come close to each other. As a consequence, there exists a critical angular momentum, $L_c$, with a corresponding critical radius $r_c$. For $L > L_c$ two circular orbits are possible: one at $r > r_c$ that is stable and the other at $r < r_c$ that is unstable. This critical behavior is analyzed as a function of the charge and the size ratios of the two spheres.
\end{abstract}

\maketitle

\section{Introduction}

Although Coulomb's law mimics Newton's law of gravitation in its $1/r^2$ behavior, no electrostatic orbits are found in nature while gravitational orbits are abundant. The reason is that electrostatic forces are much stronger than gravitational forces and form bound systems at much smaller, atomic length scales. Electrons orbit the nucleus in Bohr's atomic model, but quantum mechanics shows that atomic orbitals are much more complex.

Creating macroscopic electrostatic orbits is an experimental challenge because of the requirements of high voltages and nearly zero gravity. In addition, there are practical difficulties such as leakage of charge from the spheres to ambient air and sparking to nearby grounded objects. In August 2006 Rhodes College undergraduates successfully demonstrated electrostatic orbits\cite{TPTpaper} by making a small graphite coated negatively charged Styrofoam sphere orbit a larger stationary aluminum sphere in conditions of microgravity created aboard the NASA McDonnell Douglas C9B aircraft. In July 2008 Rhodes College undergraduates created binary orbits between two oppositely charged equal sized graphite coated Styrofoam spheres.\cite{binary}

In these experiments the spheres crash into each other if they come too close. The reason is the redistribution of charge on the two spheres due to polarization which causes the force between the two spheres to deviate significantly at shorter distances from $1/r^2$ behavior. At some separation $r$ the resultant force is able to overcome the ``centrifugal repulsion'' due to the angular momentum of the orbiting objects thus making the orbits unstable. In this paper we analyze the stability of electrostatic orbits taking polarization effects into account.

The electrostatic force between two conducting spheres with arbitrary charges and radii is a fundamental problem that has generated considerable interest.\cite{slisko} Classic books on electromagnetism by Maxwell,\cite{maxwell} Jeans,\cite{jeans} and Smythe\cite{smythe} include a detailed analysis of this problem. Maxwell credits Poisson~\cite{maxwell,maxwell2} as the first to solve the problem and Thompson (Lord Kelvin) for a ``greatly simplified'' treatment.\cite{maxwell2,thompson} More recently, Soules\cite{soules} did a numerical analysis of the force between two oppositely charged spheres of unequal sizes to show that polarization effects dominate at short distances.

The main idea behind the solution to this problem is the method of images. A complete analysis of the problem requires an infinite number of image charges. However, as shown numerically in Refs.~\onlinecite{slisko} and \onlinecite{soules}, only a finite number of image charges are needed to compute the force to a desired degree of accuracy. This number increases rapidly as the spheres come closer. For the analysis in this paper we take $120$ image charges on each sphere into account.

Because the interaction between the two spheres depends only on their center-to-center distance $r$, we can incorporate the angular motion of the objects into a centrifugal term, and reduce the problem to a one-dimensional problem.\cite{goldstein,marion} For interaction forces of the type $F(r)\sim 1/r^n$, stable orbits are possible for $n < 3$. For $n > 3$ the $1/r^3$ repulsion is unable to overcome the $1/r^n$ attraction at short distances; at large distances the repulsion becomes stronger than the attraction, and thus the stability is lost at both ends.

For our problem stability is not an issue at large distances because the Coulomb force dominates the polarization effects for large $r$. At small $r$ polarization effects become significant and higher and higher powers of $1/r$ are needed to describe the interaction accurately. In Sec.~\ref{pointcharge} we analyze the stability of a point charge orbiting a charged conducting sphere, and in Sec.~\ref{unequal} we study the more general case of two spheres with arbitrary charges and radii. We ignore energy loss due to radiation from the centripetal acceleration of the orbiting charge. By using Larmor's formula\cite{griffiths} it can be shown that the lifetime of the system due to radiation loss will be much larger than its orbital period.\cite{larger}

\section{Point charge orbiting a conducting sphere of charge}
\label{pointcharge}

Consider a conducting sphere $A$ with charge $Q$ and radius $a$ with a point charge $-q$ of mass $m$ orbiting around it. Without loss of generality we let $A$ be fixed and centered at the origin. Binary orbits between two bodies about their center of mass can be reduced to an equivalent one body problem by switching to relative coordinates.\cite{goldstein,marion} This problem requires only one image charge; it is exactly soluble and helps to understand the qualitative features of the more general case discussed in Sec.~\ref{unequal}.

Let the distance of $-q$ from the origin be $r$. Due to the presence of $-q$, the charge on $A$ is redistributed and can be considered as being composed of a charge $Q_0$ at $x_0=0$ and $Q_1 = a q /r$ at $x_1 = a^2/r$, where $x_0$ and $x_1$ are measured from the origin along the line joining the two charges. Because $Q$ is fixed, we have $Q_0 = Q - a q/r$.

\begin{figure}[h]
\begin{center}
{\includegraphics[width=1.0\linewidth]{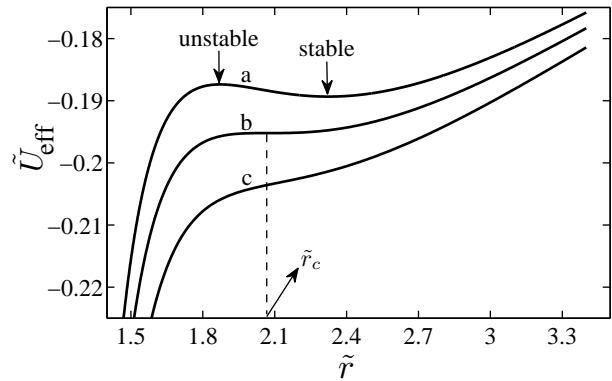}}
\end{center}
\caption{\label{fig1}$\tilde{U}_{\rm eff}(\tilde{r})$ versus $\tilde r$ for (a) $\tilde{L}^2=2.83$. The arrows indicate stable and unstable circular orbits at $\tilde{r}=2.32$ and $\tilde{r}=1.87$ respectively. (b) The critical angular momentum $\tilde{L}_c^2=2.7715$. The corresponding critical radius (indicated by the vertical dashed line) is $\tilde{r}_c=2.0704$. (c) $\tilde{L}^2=2.70$ with no possible orbits.}
\end{figure}

If we use the principle of superposition, the net force on the point charge can be written as
\begin{subequations}
\label{force1}
\begin{align}
F_{E}(r) &= - k q \left({Q_0 \over r^2} + {Q_1 \over (r-x_1)^2}\right), \\
&= - k q \left({Q \over r^2} -{a q\over r^3}+ {{a q r} \over {(r^2-a^2 )^2}}\right ),
\end{align}
\end{subequations}
where $k = 1/4 \pi \epsilon_0$.
Integrating Eq.~(\ref{force1}) with respect to $r$ (from $\infty$ to $r$) gives the
electrostatic interaction energy of the system as
\begin{equation}
\label{potential1}
U_{E}(r) = - k q \bigg({Q \over r} -{q a \over 2r^2}+ {q a \over 2(r^2-a^2)}\bigg).
\end{equation}
The effective potential energy for the system is\cite{goldstein,marion}
\begin{equation}
\label{effpotential}
U_{\rm eff}(r)= {L^2 \over {2m r^2}}+ U_{E}(r),
\end{equation}
where $L$ is the total angular momentum of the system, and the first term in the energy is the centrifugal term.
For numerical and analytical calculations it is convenient to define the dimensionless variables
\begin{subequations}
\label{dimensionless}
\begin{align}
\tilde{r} & \equiv {r \over a }, \quad \tilde{Q} \equiv {Q \over q }, \quad \tilde{L}^2 \equiv {L^2 \over k q^2 a m},\\
\tilde{U}_{E} & \equiv {U_{E} a \over k q^2},\ \mbox{and}\ \tilde{U}_{\rm eff} \equiv {U_{\rm eff} a \over k q^2}.
\end{align}
\end{subequations}
We first set $\tilde{Q}=1$ and plot $\tilde{U}_{\rm eff}(\tilde{r})$ for several values of $\tilde{L}^2$. For $\tilde{L}^2 = 2.83$ (see curve (a) in Fig.~\ref{fig1}), the effective potential has two extremum points corresponding to two circular orbits: the one at $\tilde{r}=2.32$ is stable (${d^2\tilde{U}_{\rm eff}/d\tilde{r}^2}> 0$), and the one at $\tilde{r}=1.87$ is unstable (${d^2\tilde{U}_{\rm eff}/d\tilde{r}^2}<0$). This smaller unstable orbit is purely due to polarization effects because at sufficiently close distances the polarization effects are so strong that they overcome the centrifugal repulsion. As we decrease the value of $\tilde{L}^2$ we find that there exists a critical value $\tilde{L_c}^2 = 2.7715$ where the two extrema merge into a single point at the critical radius, $\tilde{r}_c = 2.0704$ (see curve (b) in Fig.~\ref{fig1}). There are no stable or unstable orbits below this critical value of $\tilde{L}^2$ (see curve (c) in Fig.~\ref{fig1}).

To obtain the critical parameters we set
\begin{subequations}
\label{criteria}
\begin{align}
{d\tilde{U}_{\rm eff} \over d\tilde{r}}& =0 \label{criteria.a}\\
\noalign{\noindent and}
{d^2\tilde{U}_{\rm eff} \over d\tilde{r}^2}& =0. \label{criteria.b}
\end{align}
\end{subequations}
By eliminating $\tilde{L}^2$ from these conditions, we can write the equation for $\tilde{r}_c$ as
\begin{equation}
\label{rc}
\left[ 3 {d\tilde{U}_E \over d\tilde{r}} + \tilde{r} {d^2\tilde{U}_E \over d\tilde{r}^2}\right]_{\tilde{r}=\tilde{r}_c}=0.
\end{equation}
The substitution of $\tilde{U}_E$ (defined through Eqs.~(\ref{potential1}) and (\ref{dimensionless})) into Eq.~(\ref{rc}) gives
\begin{equation}
\label{rc2}
\left[{\tilde{Q} \over \tilde{r}^2} = {4 \tilde{r} \over (\tilde{r}^2 -1)^3}\right]_{\tilde{r}=\tilde{r}_c}.
\end{equation}
On the left-hand side of Eq.~(\ref{rc2}) is the Coulomb force due to the first term in $U_E$ from Eq.~(\ref{potential1}), and on the right-hand side is the expression for the polarization force.\cite{footnote1} The point at which these two are equal determines the critical radius.
Note that the polarization force is independent of $\tilde{Q}$ because the charge on the sphere is unable to polarize the point charge. It goes to $\infty$ in the limit $\tilde{r} \to 1$ as expected. For large $\tilde{r}$ it goes as $1/\tilde{r}^5$ because the orbiting point charge creates a dipole of order $1/\tilde{r}^2$ in sphere $A$ that interacts with the point charge with its dipolar field that goes as $1/\tilde{r}^3$.

The solution of Eq.~(\ref{rc2}) for $\tilde{r}_c$ gives
\begin{equation}
\label{rcresult}
\tilde{r}_c = {1 + \sqrt {1 + z^2} \over z},
\end{equation}
where $z={(2 \tilde{Q})}^{1/3}$. In the limit $\tilde{Q} \to \infty$ Eq.~(\ref{rcresult}) gives $\tilde{r}_c = 1$ (or $r_c = a$) because the Coulomb force in Eq.~(\ref{rc2}) dominates the polarization force for $\tilde{r}>1$. In the other limit $\tilde{Q} \to 0$, $\tilde{r}_c \to \infty$ as $1/\tilde{Q}^{1/3}$. The polarization force dominates the Coulomb force for $\tilde{r} \neq 0$ in this limit, and Eq.~(\ref{rc2}) in the leading order becomes ${\tilde{Q}/\tilde{r}_c^2} = 4/\tilde{r}_c^5$. Solving for $\tilde{r}_c$ gives the $1/\tilde{Q}^{1/3}$ behavior.

The substitution of $\tilde{r}_c$ in Eq.~(\ref{criteria.a}) or Eq.~(\ref{criteria.b}) gives
\begin{equation}
\label{lcresult}
\tilde{L}_c^2 = z^2~ {12 + 7 z^2 +(12 + 2 z^2)\sqrt {1 + z^2} \over 4\left( 1+ \sqrt {1 + z^2}\right)^2}.
\end{equation}
In the limit $\tilde{Q} \to \infty$, $\tilde{L}_c^2$ approaches $\infty$ as the first power of $\tilde{Q}$. This result can be understood by setting the centrifugal repulsion $\tilde{L}_c^2/\tilde{r}_c^3$ equal to the Coulomb force\cite{footnote2} and noting that $\tilde{r}_c = 1$. For $\tilde{Q}\to 0$, $\tilde{L}_c^2$ approaches $0$ as $\tilde{Q}^{2/3}$. This result follows from the same argument as before and noting that $\tilde{r}_c \sim 1/\tilde{Q}^{1/3}$ in this limit.

\section{Orbiting charged conducting spheres of unequal size}
\label{unequal}
We now consider the more general case of two conducting spheres $A$ and $B$ with charges $q_1=Q$ and $q_2=-q$ and radii $a$ and $b$, respectively. To calculate the effective potential energy in Eq.~(\ref{effpotential}) we need the electrostatic interaction energy between the two spheres. As noted by Maxwell, this calculation requires ``an intricate investigation''\cite{maxwell2} because the two spheres create an infinite series of image charges on themselves and on each other.

The total electrostatic energy of $A$ and $B$ (including their self energies) is given by\cite{maxwell,jeans}
\begin{equation}
\label{energy1}
W_V(r)= {1\over 2}( c_{11} V_1^2 + 2 c_{12} V_1 V_2 + c_{22} V_2^2),
\end{equation}
where $V_1$ and $V_2$ are the voltages of $A$ and $B$ respectively, and the coefficients $c_{ij}$ are their self and mutual capacitances defined by the relations
\begin{subequations}
\label{capacitancedef}
\begin{align}
q_1 &= c_{11} V_1 + c_{12} V_2 \\
q_2 &= c_{21} V_1 + c_{22} V_2.
\end{align}
\end{subequations}
We use these relations to express the energy in Eq.~(\ref{energy1}) in terms of the charges of the spheres as
\begin{equation}
\label{energy2}
W_q(r)= {1\over 2}( p_{11} q_1^2 + 2 p_{12} q_1 q_2 + p_{22} q_2^2),
\end{equation}
where the matrix $p_{ij}$ is the inverse of the $2 \times 2$ capacitance matrix $c_{ij}$.

To calculate the coefficients $c_{ij}$ the interaction between the spheres is ignored as a zeroth order approximation, giving a charge of $a V_1/k$ on $A$ and $b V_2/k$ on $B$. We then use the method of images to calculate successive corrections to the charges on each sphere so that they remain fixed at their given potentials $V_1$ and $V_2$. The infinite series of charges on the spheres is summed and set equal to their net charges $q_1$ and $q_2$ and the coefficients of capacitance are identified from Eq.~(\ref{capacitancedef}). Poisson and Kirchoff\cite{maxwell, jeans} calculated these coefficients and expressed them as
\begin{subequations}
\label{capacitance}
\begin{align}
c_{11} &= {a \over k}(1 - \xi^2) \sum_{n=0}^\infty {\alpha^n \over 1 - \xi^2 \alpha^{2n}},\\
c_{12} &= c_{21} = -{a b \over k r} (1 - \alpha^2) \sum_{n=0}^\infty {\alpha^n \over 1 - \alpha^{2n-2}},\\
c_{22} &= {b \over k} (1 - \eta^2) \sum_{n=0}^\infty {\alpha^n \over 1 - \eta^2 \alpha^{2n}},
\end{align}
\end{subequations}
where
\begin{equation}
\xi^2 = {(a+ b \alpha)^2 \over r^2}~~~~\mbox{and}~~~~\eta^2 = {(b+ a \alpha)^2 \over r^2},
\end{equation}
and $\alpha$ is the smaller root of the quadratic equation
\begin{equation}
t^2 - t {(r^2 - a^2 -b^2) \over {a b}} + 1 = 0.
\end{equation}
In the limit $r \to \infty$, $c_{11} = a/k$, $c_{22} = b/k$, and $c_{12} = c_{21} = 0$ as expected. We truncate the series for $c_{ij}$ in Eq.~(\ref{capacitance}) at $n=60$. This truncation takes into account the contribution of 120 image charges on each sphere in addition to the charges at their centers.

From the capacitance matrix $c_{ij}$ we calculate its inverse matrix $p_{ij}$ and substitute it into Eq.~(\ref{energy2}) and expand in powers of $1/r$ to obtain the total electrostatic energy (divided by $k$) as
\begin{eqnarray}
\label{Wqexpansion}
{W_q(r) \over k} &=& {q_1^2 \over 2 a} + {q_2^2 \over 2 b} - {q_1 q_2 \over r} - {b^3 q_1^2 + a^3 q_2^2 \over 2 r^4} - {b^5 q_1^2+ a^5 q_2^2 \over 2 r^6} \nonumber \\
&&{}- {2 a^3 b^3 q_1 q_2 \over r^7} - {b^7 q_1^2 + a^7 q_2^2 \over 2 r^8} + \ldots
\end{eqnarray}
After subtracting the self energies $q_1^2/ 2a$ and $q_2^2/ 2b$ of $A$ and $B$, and then setting $q_1 = Q$ and $q_2=-q$, we expand their electrostatic interaction energy (divided by $k q^2/a$) in powers of $1/\tilde{r}$ to obtain
\begin{eqnarray}
\label{expansion2}
\tilde{U}_E (\tilde{r})&=& - {\tilde{Q} \over \tilde{r}} - {s^3 \tilde{Q}^2 + 1 \over 2 \tilde{r}^4} - {s^5 \tilde{Q}^2+1 \over 2 \tilde{r}^6}- {2 s^3 \tilde{Q} \over \tilde{r}^7} \nonumber
\\
&&{}- {s^7 \tilde{Q}^2+1 \over 2 \tilde{r}^8} - {3 (s^3 + s^5) \tilde{Q} \over \tilde{r}^9} + \ldots
\end{eqnarray}
where $s \equiv b/a$ is the size ratio of the spheres. This form of the interaction energy lets us easily calculate the derivatives and double derivatives needed in the following. Including $120$ image charges on each sphere allows us to carry out this expansion to order $(1/\tilde{r})^{121}$ and gives numerical values for the force ($-d\tilde{U}_E/d\tilde{r}$) accurate to $\epsilon < 10^{-6}$ for $r \ge 1.1 (a+b)$ and $ 0.01 \le \tilde{Q} \le 100$. The quantity $\epsilon$ is the fractional change in the force due to the inclusion of one additional term in the approximation of $c_{ij}$ in Eq.~(\ref{capacitance}).

The rest of the treatment mirrors that in Sec.~\ref{pointcharge}. Without loss of generality we assume $s \le 1$ and let $A$ be fixed at the origin and $B$ be the orbiting sphere. The effective potential energy is defined using Eqs.~(\ref{effpotential}) and (\ref{dimensionless}). To find the critical radius we
substitute $\tilde{U}_E$ from Eq.~(\ref{expansion2}) into Eq.~(\ref{rc}) and obtain
\begin{eqnarray}
&&\bigg [{\tilde{Q} \over \tilde{r}^2} = {4(s^3 \tilde{Q}^2 + 1 )\over \tilde{r}^5} + {12(s^5 \tilde{Q}^2+1) \over \tilde{r}^7}+ {70 s^3\tilde{Q} \over \tilde{r}^8} \nonumber \\
&&{} \quad + {24(s^7 \tilde{Q}^2+1) \over \tilde{r}^9} + {189 (s^3 + s^5) \tilde{Q} \over \tilde{r}^{10}}+ \ldots {\bigg]}_{\tilde{r}=\tilde{r}_c}.\label{rc3}
\end{eqnarray}
Equation~\eqref{rc3} is the general form of Eq.~(\ref{rc2}) and takes into account the finite size of the orbiting sphere. At the critical radius the Coulomb force on the left has to balance the infinite series for the polarization force on the right. We solve Eq.~({\ref{rc3}) numerically for $\tilde{r}_c$ as a function of the charge ratio $\tilde{Q}$ for several values of the size ratio $s$. The results are plotted in Fig.~\ref{rcvsQ}.
\begin{figure}[h]
\begin{center}
{\includegraphics[width=1.0\linewidth]{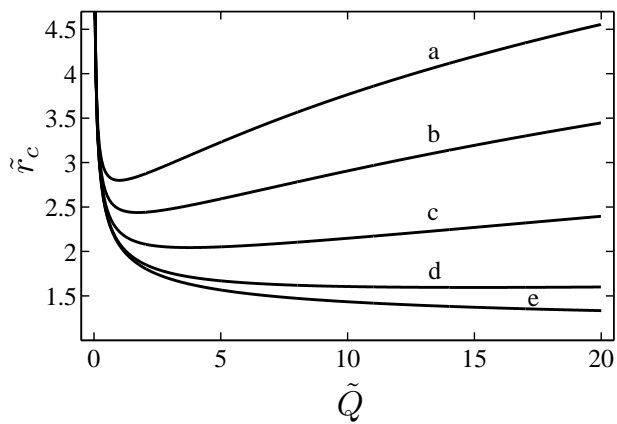}}
\end{center}
\caption{\label{rcvsQ}The critical radius, $\tilde{r}_c$, plotted against the charge ratio $\tilde{Q}$ for size ratios of (a) $s = 1$, (b) $s = 3/4$, (c) $s = 1/2$, (d) $s = 1/4$, and (e) $s = 0$. All curves with non-zero size ratio asymptotically converge to the $s=0$ curve which goes as $1/\tilde{Q}^{1/3}$ in the limit $\tilde{Q}\rightarrow 0$. In the limit $\tilde{Q}\rightarrow \infty$ they go as $s \tilde{Q}^{2/3}$ except for the $s=0$ case which goes asymptotically to $1$.}
\end{figure}

For $\tilde{Q} \to 0$ Eq.~({\ref{rc3}) becomes independent of $s$, and we obtain the $s=0$ case discussed in Sec.~\ref{pointcharge}.
In this limit all the $\tilde{r}_c$ versus $\tilde{Q}$ curves asymptotically approach the $s=0$ curve, which goes to $\infty$ as $1/\tilde{Q}^{1/3}$. For $\tilde{Q} \to \infty$, due to the $\tilde{Q}^2$ terms in Eq.~({\ref{rc3}), the polarization force dominates the Coulomb force for any finite value of $\tilde{r}$. The equation for the critical radius in this limit is $ {\tilde{Q}/\tilde{r}_c^2} = {4 s^3 \tilde{Q}^2/\tilde{r}_c^5}$. Thus, $\tilde{r}_c$ goes to $\infty$ as $s \tilde{Q}^{1/3}$ as seen in Fig.~\ref{rcvsQ}. Note that for all $s\ne 0$, the critical radius has a minimum, $\tilde{r}_{c,\min}$, with respect to $\tilde{Q}$. This minimum is expected because $\tilde{r}_c$ goes to $\infty$ at both ends of the $(0,\infty)$ domain for $\tilde{Q}$. The numerical values of $\tilde{r}_{c,\min}$ are listed in Table~\ref{closestapproach} along with the corresponding values of $\tilde{L}_{c,min}^2$ and the charge ratio $\tilde{Q}_{\min}$ at which they occur.
\begin{table}[h]
\begin{center}
\begin{tabular}{|c| c| c| c|}
\hline
$s$& $\tilde{Q}_{\min}$ & $\tilde{r}_{c,\min}$& $\tilde{L}_{c,\min}^2$  \\
\hline \hline
1 & 1 & 2.800 & 3.603 \\
\hline
3/4 & 1.725 & 2.438 & 5.405  \\
\hline
1/2 & 3.760 & 2.044 & 9.798 \\
\hline
1/4 & 15.11 & 1.595 & 29.93  \\
\hline
1/8 & 66.92 & 1.336 & 106.8 \\
\hline
0 & $\infty$ & 1 & $\infty$ \\
\hline
\end{tabular}
\caption{\label{closestapproach} The values of the charge ratio, $\tilde{Q}_{\min}$, at which the minimum critical radius occurs are listed (see second column) versus the size ratio $s$. The third column gives the value of this minimum critical radius and the fourth column gives the corresponding value of $\tilde{L}_{c,\min}^2$. For smaller size ratios the fixed sphere can hold a greater charge than the orbiting sphere, and the limiting orbits are smaller and faster as indicated by the last two columns.}
\end{center}
\end{table}

For $s= 1$ the minimum critical radius is at $\tilde{Q}_{\min}=1$ as expected, because the result has to be symmetric with respect to the inversion $\tilde{Q} \to 1/\tilde{Q}$ ($q \leftrightarrow Q$). Equation~(\ref{rc3}) is invariant under this transformation as be seen by dividing by $\tilde{Q}$ on both sides. For $s < 1$ the $\tilde{Q} \to 1/\tilde{Q}$ symmetry is broken. As $s$ becomes smaller, the minimum critical radius decreases and the value of $\tilde{Q}_{\min}$ at which it occurs increases (see Table~\ref{closestapproach} and Fig.~\ref{rcvsQ}). Qualitatively, as the size ratio decreases, the larger fixed sphere is able to hold a greater charge without excessively polarizing the smaller orbiting sphere. This leads to smaller stable orbits with higher angular momentum.

The numerical solution of $\tilde{r}_c$ can be substituted into Eqs.~(\ref{criteria.a}) or (\ref{criteria.b}) to obtain the corresponding solution of $\tilde{L}_c^2$. The results are plotted in Fig.~\ref{L2vsQ}. In the limit $\tilde{Q} \to 0$, Eq.~({\ref{rc3}) is independent of $s$, and all the curves approach zero as $\tilde{Q}^{2/3}$, which is the same behavior as found for a point charge. The $\tilde{Q} \to \infty$ limit can be obtained by setting the centrifugal force $\tilde{L}_c^2/\tilde{r}_c^3$ equal to the Coulomb force ${\tilde{Q}/\tilde{r}_c^2}$. In this limit $\tilde{r}_c \sim s\tilde{Q}^{1/3}$ and thus $\tilde{L}_c^2 \to \infty$ as $s\tilde{Q}^{4/3}$.
\begin{figure}[h]
\begin{center}
{\includegraphics[width=1.0\linewidth]{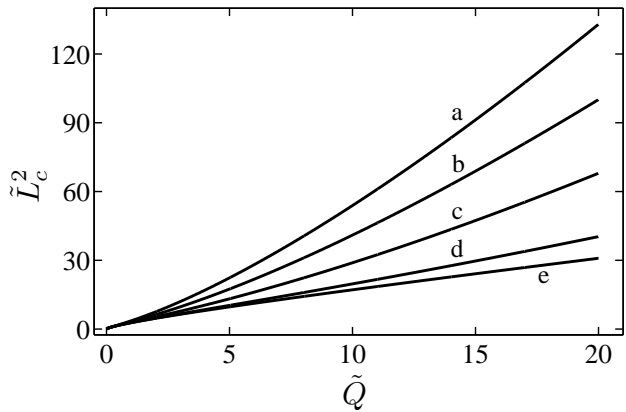}}
\end{center}
\caption{\label{L2vsQ}The critical angular momentum, $\tilde{L}_c^2$, plotted against the charge ratio $\tilde{Q}$ for size ratios of (a) $s = 1$, (b) $s = 3/4$, (c) $s = 1/2$, (d) $s = 1/4$, and (e) $s = 0$. All the curves approach zero as $\tilde{Q}^{2/3}$ independent of the size ratio. In the limit $\tilde{Q}\rightarrow \infty$ they go as $s\tilde{Q}^{4/3}$ except for the $s=0$ case which is linear in $\tilde{Q}$ in that limit.}
\end{figure}

It it interesting to compare this theory to the successful orbits reported in Table I of Ref.~{\onlinecite{TPTpaper}}. The size ratio of the spheres for these experiments was $s=1.6\,\mbox{cm}/6.5\,\mbox{cm} = 0.25$, and the charge ratio, $\tilde{Q}$, (estimated from the given voltages\cite{estimate}) ranged from $5.0$ to $5.8$. For these values the critical parameter $\tilde{L}_c^2$ is between $10.3$ and $11.8$ and $\tilde{r}_c$ is $1.7$. The successful orbits had initial values\cite{velocity} of $13 \le \tilde{L}^2 \le 42$ and $2.2 \le \tilde{r} \le 2.8$, which are above the threshold needed for stability.

\section{Conclusions}

We analyzed the stability of orbits between two charged conducting spheres due to their electrostatic attraction. For a two body gravitational system only one stable circular orbit is possible for a given value of angular momentum. In contrast, for an electrostatic system two circular orbits are possible for a fixed angular momentum. The larger of these two orbits is stable and the smaller orbit is unstable.

We also found that there exists a critical angular momentum below which no stable or unstable orbits are possible. The critical radius corresponding to this critical angular momentum is the limiting value of the smallest possible stable and largest possible unstable orbits. The critical parameters in dimensionless form depend only on the charge and the size ratio of the two spheres.

Analyzing the critical parameters with respect to the charge and size ratios yields a minimum critical radius for every non-zero value of the size ratio. This minimum critical radius occurs at the charge ratio of unity when the spheres are equal in size. As the size ratio decreases, the charge ratio where the minimum critical radius occurs increases while the value of the minimum critical radius decreases. In the limit of zero size ratio (point charge), the charge ratio approaches $\infty$ and the minimum critical radius approaches the radius of the stationary sphere.

We suggest two possible undergraduate research projects as follow ups to the work in this paper. Although zero gravity conditions are ideal for observing the stability of electrostatic orbits, ground based experiments can be performed by hanging the orbiting sphere from a high ceiling. The orbits can also be simulated using the equations of motion and the full electrostatic interaction (including polarization effects) between the two spheres.

\begin{acknowledgments}
We thank Dr.\ Brent Hoffmeister for useful discussions and Dr.\ Charles Roberston for funding the summer support of the undergraduate student.
\end{acknowledgments}


\begin{thebibliography}{8}

\bibitem{TPTpaper} S. Banerjee, K. W. Andring, D. L. Campbell, J. A. Janeski, D. A. Keedy, S. P. Quinn, and B. K. Hoffmeister, ``Orbital motion of electrically charged spheres in microgravity," Phys. Teach. {\bf 46}, 460--464 (2008).

\bibitem{binary} The results are yet to be published.

\bibitem{slisko} See J. Sli\v{s}ko and R. A. Brito-Orta, ``On approximate formulas for the electrostatic force between conducting spheres," Am. J. Phys. {\bf 66}, 352--355 (1998) and references therein.

\bibitem{maxwell} J. C. Maxwell, {\it A Treatise on Electricity and Magnetism} (Dover, New York, 1954), Vol. 1. pp. 281--283.

\bibitem{jeans} J. Jeans, {\it The Mathematical Theory of Electricity and Magnetism} (Cambridge University Press, Cambridge, 1966), 5th ed. pp. 196--199.

\bibitem{smythe} W. R. Smythe, {\it Static and Dynamic Electricity} (McGraw-Hill, New York, 1968), 3rd ed., pp. 224--231.

\bibitem{maxwell2} Reference~\onlinecite{maxwell}, p. 47.

\bibitem{thompson} W. Thompson, ``On the mutual attraction or repulsion between two electrified spherical conductors,'' in W. Thompson, {\it Reprint of Papers on Electrostatics and Magnetism} (Mcmillan, London, 1872), pp. 86--97.

\bibitem{soules} J. A. Soules, ``Precise calculation of the electrostatic force between charged spheres including induction effects," Am. J. Phys. {\bf 58}, 1195--1199 (1990).

\bibitem{goldstein} H. Goldstein, C. P. Poole, and J. L. Safko, {\it Classical Mechanics}
(Addison-Wesley, New York, 2002) 3rd ed., pp. 70--102.

\bibitem{marion} J. B. Marion and S. T. Thornton, {\it Classical Dynamics of Particles and Systems} (Saunders, San Diego, 1995), 4th ed., pp. 291--317.

\bibitem{griffiths} D. J. Griffiths, {\it Introduction to Electrodynamics}, (Prentice Hall, Upper Saddle River, 1999), 3rd ed., p. 458.

\bibitem{larger} The ratio of the lifetime to the orbital period is of order $c^3/v^3$, where $v$ is the orbital speed and $c$ is the speed of light.

\bibitem{footnote1} The polarization force, defined as the right hand side of Eq.~(\ref{rc2}), comes from substituting the non-Coulombic terms in Eq.~(\ref{potential1}) into Eq.~(\ref{rc}). It is not equal to the total force minus the Coulomb force.

\bibitem{footnote2} At the critical point the centrifugal force is of the same order as but is not exactly equal to the Coulomb force.

\bibitem{estimate} Ground based tests showed that the orbiting sphere received only about $80\%$ of the maximum possible charge at that voltage due to the small but finite size of the electrode which ``shared'' the charge with the sphere.

\bibitem{velocity} In addition to the parameters listed in Table I of Ref.~\onlinecite{TPTpaper} the launch velocities obtained from the video of the orbits were used in calculating the values of $\tilde{L}^2$. A link to the video is provided in Ref.~\onlinecite{TPTpaper}.

\end{thebibliography}
\end{document}